\documentclass[ceurart]{ceurart}

\usepackage{libertinus}
\usepackage{listings}
\usepackage{xcolor}
\begin{document}

\copyrightyear{2025}
\copyrightclause{Copyright for this paper by its authors.
  Use permitted under Creative Commons License Attribution 4.0
  International (CC BY 4.0).}


\conference{Proceedings of RDGENAI '25: First International Workshop on Retrieval-Driven Generative AI at CIKM 2025, November 14th, 2025, Coex, Seoul, Republic of Korea}

\title{MARC: Multimodal and Multi-Task Agentic Retrieval-Augmented Generation for Cold-Start Recommender System}

\tnotemark[1]

\author[1]{Seung Hwan Cho}[%
email=shcho95@hayang.ac.kr,
]
\fnmark[1]

\author[1]{Yujin Yang}[%
email=yujinyang@hanyang.ac.kr,
]
\fnmark[1]

\author[1]{Danik Baeck}[%
]

\author[1]{Minjoo Kim}[%
]

\author[1, 2]{Young-Min Kim}[%
orcid=0000-0002-6914-901X,
email=yngmnkim@hanyang.ac.kr,
]
\cormark[1]

\author[1, 2]{Heejung Lee}[%
]

\author[1, 2]{Sangjin Park}[%
]

\address[1]{Department of Industrial Data Engineering, Hanyang University, Republic of Korea}
\address[2]{School of Interdisciplinary Industrial Studies, Hanyang University, Republic of Korea}

\cortext[1]{Corresponding author.}
\fntext[1]{These authors contributed equally.}

\begin{abstract}
Recommender systems (RS) are currently being studied to mitigate limitations during the cold-start conditions, by leveraging modality information or introducing Agent concepts based on the exceptional reasoning capabilities of Large Language Models (LLMs). Meanwhile, food and beverage recommender systems have traditionally used knowledge graph and ontology concepts due to the domain's unique data attribute and relationship characteristics. On this background, we propose MARC, a multimodal and multi-task cocktail recommender system based on Agentic Retrieval-Augmented Generation (RAG) utilizing graph database under cold-start conditions. The proposed system generates high-quality, contextually appropriate answers through two core processes: a task recognition router and a reflection process. The graph database was constructed by processing cocktail data from Kaggle, and its effectiveness was evaluated using 200 manually crafted questions. The evaluation used both LLM-as-a-Judge and human evaluation to demonstrate that answers generated via the graph database outperformed those from a simple vector database in terms of quality. The code is available at https://github.com/diddbwls/cocktail\_rec\_agentrag
\end{abstract}
	

\maketitle

\section{Introduction}
In Natural Language Processing (NLP), Large Language Models (LLMs) are shifting research paradigms through their exceptional reasoning capabilities \cite{Brown:2020aa, Zhao:2023aa}. LLMs demonstrate outstanding performance across diverse domains and tasks when combined with methodologies such as prompt engineering \cite{Radford:2019aa}, fine-tuning \cite{Hu:2022aa}, and Retrieval-Augmented Generation (RAG) \cite{Lewis:2020aa}. Recent research in Recommender Systems (RS) have also leveraged these LLMs capabilities \cite{Wang:2023aa, Lyu:2023aa, Shi:2024aa, Tian:2024aa}, with studies introducing the concept of Agent being introduced \cite{Wang:2023ab, Shu:2024aa, Huang:2025aa}. It is important to note that the Agent concept is utilized extensively without clear boundaries in its definition \cite{Bent:2025aa}. This ambiguity can cause confusion in understanding research content. Therefore, this study bases its concepts on the definitions of Agent and Agentic RAG established by Wang et al. \cite{Wang:2024aa} and Singh et al. \cite{Singh:2025aa}.

Recommender systems are information retrieval mechanisms that facilitate decision-making by providing internet users with information regarding their preferences and interests, in cases where the users lack the necessary information or expertise \cite{Wang:2023ac}. Following the widespread adoption of the internet, a substantial volume of diverse data has been generated online, thereby presenting users with the challenge of rapidly identifying the information they require. Consequently, recommender systems have evolved to provide users with suitable content. Starting with recommender systems applying collaborative filtering methods between users \cite{Resnick:1994aa}, research has actively pursued performance enhancement from various perspectives, including modeling, scalability, and sequentiality \cite{Sarwar:2001aa, Koren:2009aa, Covington:2016aa}. These collaborative filtering-based methodologies have been shown to demonstrate excellent performance in warm-start characterized by abundant user-item interactions. Conversely, they have been observed to struggle to guarantee performance in cold-start with sparse interactions \cite{Volkovs:2017aa}. To mitigate this limitation, content-based research utilizing modality information such as text, images, and tables has been conducted \cite{Li:2023aa, Yuan:2023aa}. 

In application domains of recommender systems where data structures, attributes and relationships are critical, knowledge graph or ontology based methodologies are employed. A representative case is food and beverage recommender systems, where these approaches are used to accurately capture the intricate relationships between ingredients, recipes, containers, and other factors \cite{Haussmann:2019aa, Oliveira:2021aa, Gawrysiak:2024aa}. This suggests that while user interaction and history are important, meaningful recommendations can be made based solely on semantic similarity or relational information between content items. Moreover, knowledge graph has the advantage of being able to automatically explain the process of deriving answers, thereby enabling user understanding and reuse \cite{Haussmann:2019aa}.

In this paper, we propose MARC, a multimodal and multi-task cocktail recommender system based on Agentic RAG using a graph database under cold-start conditions, aligned with recent research trends and domain specificity. The graph database used for RAG is cocktail data, which is based on the relationships between features. The proposed system has two main stages: Task Recognition Router and Reflection. Starting with the user's image and text query (i.e., multimodal), the Router determines the task and performs retrieval through a configured algorithm based on this determination. The retrieved information is then evaluated by LLMs to verify that the data has been correctly fetched from and generated by the graph database. This Reflection continues until a configured threshold is exceeded. The resulting context from the Reflection is then fed into the model alongside the user's query and prompt template. This enables the model to generate recommendations in response to the query. The main contributions of our study are as follows:

\begin{itemize}
\item We propose an Agentic RAG based recommender system that integrates multimodal and multi-task LLMs with Graph RAG for recommendations on cocktail data.
\item We designed a Task Recognition Router and Reflection mechanism enabling multimodal and multi-task recommendations, thereby improving answer generation quality without requiring special fine-tuning.
\item We demonstrate the effectiveness of the proposed method in the cocktail domain and confirm its potential as an interactive and explainable beverage recommender system.
\end{itemize}

\section{Related Works}

\subsection{Beverage and Food Recommendation}
Recommender systems based on food or beverages are not mainstream research compared to the movie domain, but they are steadily being studied, with research utilizing ontologies or knowledge graphs existing. Ontologies and knowledge graphs are representations that systematize information and relational data into a graph structure based on a knowledge system and schema of concept definitions and relationships through domain-specific vocabulary. They provide the ability to reflect constraints necessary for recommendation and offer explainability \cite{Lim:2004aa, Chen:2012aa, Chen:2019aa}. 

Chen et al. proposed a recommender system combining common-sense reasoning to detect emotions from inferences and colors based on a cocktail ontology knowledge base  \cite{Chen:2006aa}, while Ahlam et al. built an ontology based on the Canada Food Allergies and Intolerances Databases and integrated useful features such as food item selection, descriptions, and recommendations \cite{Al:2019aa}. Haussmann et al. constructed a knowledge graph based on large-scale public food data and performed food recommendations \cite{Haussmann:2019aa}. Oliveira et al. constructed a wine ontology and confirmed that ontologies in recommender systems influenced performance improvement  \cite{Oliveira:2021aa}. Showafah et al. constructed an ontology containing knowledge about food and nutritional components along with nutritional requirements, and combined it with the TOPSIS method to provide optimal recommendations regarding nutritional balance and user preferences \cite{Showafah:2021aa}. Chen et al. proposed a collaborative recipe knowledge graph, combining health suitability scoring between recipes and user preferences with an attention-based graph convolutional neural network to present a health-aware personalized recommendation model \cite{Chen:2023aa}. Gawrysiak et al. proposed WineGraph, which pairs food and wine using food and wine review data and augments FlavorGraph data \cite{Gawrysiak:2024aa}.

\subsection{Graph Retrieval Augmented Generation}
Graph RAG is a method that addresses the limitations of vector-based RAG. By leveraging the semantic and structural relationships between nodes in a knowledge graph structure, it simultaneously improves performance and interpretability in multi-document reasoning and relationship-based question answering through search and inference \cite{Han:2024aa}. Han et al. proposed GraphRAG, which constructs a graph from entities and relationships extracted from documents and integrates it as input for LLMs through subgraph search \cite{Han:2024aa}. Hu et al. proposed GRAG, performing efficient subgraph exploration using a divide-and-conquer approach \cite{Hu2024GRAGGR}, while Shen et al. proposed GeAR, combining an agent framework with graph expansion \cite{Shen:2024aa}. Xiang et al. demonstrated through GraphRAG-Bench that graph structures are particularly effective for questions where relational information is crucial \cite{Xiang:2025aa}. Agrawal et al. introduced query-based graph neural networks to enhance query-aware retrieval performance \cite{Agrawal:2025aa}. Liang et al. proposed a graph-based RAG specialized for the geospatial analysis domain, demonstrating domain-tailored applicability \cite{Liang:2025aa}. In summary, Graph RAG has complemented existing RAG in terms of accuracy, efficiency, and explainability, establishing itself as a core technology directly relevant to Agentic RAG design in this research.

\subsection{LLM based Recommendation}
In the early stages of recommender systems, Collaborative Filtering (CF) was applied to newsgroups to make recommendations based on user rating similarity \cite{Resnick:1994aa}, and content-based methodologies \cite{Sarwar:2001aa}, Matrix Factorization (MF) \cite{Koren:2009aa}, and deep learning-based methodologies \cite{Covington:2016aa} were introduced, leading to significant advancements. Building upon this prior research, this discussion will focus exclusively on the rapidly evolving recommender systems based on LLMs.
Wang et al. proposed a zero-shot Next-Item Recommendation (NIR) prompting strategy integrating a three-step prompting approach to recommend movie ranking lists  \cite{Wang:2023aa}. Lyu et al. proposed LLM-REC, integrating four unique prompting strategies, to enhance personalized text-based recommendations through text augmentation  \cite{Lyu:2023aa}. Tian et al. improved recommendation accuracy and relevance by processing multimodal information through LLMs and projecting it into an integrated latent space \cite{Tian:2024aa}. Finally, Wang et al., whose work is most similar to ours, aimed to enhance answer generation and recommendation quality by supplementing LLMs' hallucination issues with KG RAG \cite{Wang:2025aa}. Our research distinguishes itself by introducing an attempt to build an Agentic RAG framework that autonomously recognizes user queries and improves answer quality.
	
\subsection{LLM as Recommendation Agent}
Research utilizing LLM as agent has shifted beyond the existing static prompt-response paradigm, transforming it into a dynamic decision-making framework capable of managing complex systems through diverse sub-component configurations \cite{Patil:2024aa}. A typical LLM-based agent structure consists of four modules: Profile, Memory, Planning, and Action \cite{Wang:2025aa}. Research introducing agents based on this structural form is also underway in recommender systems, which can be categorized into recommender-oriented, interaction-oriented, and simulation-oriented approaches \cite{Peng:2025aa}.

The recommender-oriented approaches focus on developing intelligent recommender systems with planning capabilities, making direct recommendations based on users' past behaviors \cite{Wang:2023ab, Shi:2024aa, Zhao:2024aa}. Wang et al. proposed a self-inspiring algorithm to enhance planning capabilities, developing an agent capable of zero-shot personalized recommendations \cite{Wang:2023ab}, while Shi et al. proposed a framework for learning planning capabilities at the macro level through reflector reflection and at the micro level through personalized recommendations via interactions between agent and critic \cite{Shi:2024aa}. Zhao et al. proposed a framework that sets LLMs as proxy users and recommends through tool learning, enabling the generation of recommendation lists aligned with preferences \cite{Zhao:2024aa}. 

Interaction-oriented approaches enable tracking user preferences and explaining recommendation rationale through conversation \cite{Shu:2024aa, Zeng:2024aa, Huang:2025aa}. Zeng et al. developed an interactive agent using LLMs and Answer Set Programming (ASP) that can request missing information \cite{Zeng:2024aa}, while Shu et al. proposed a human-centered recommendation framework based on a learn-execute-critique loop and a reflection mechanism to ensure alignment with user personality \cite{Shu:2024aa}. Huang et al. enhanced the agent component and integrated traditional recommender systems to build a multi-purpose interactive system \cite{Huang:2025aa}.

The simulation-oriented approaches aim to simulate user behavior and item characteristics within recommender systems \cite{Zhang:2024aa, Zhang:2024ab, Guo:2024aa}. Zhang et al. proposed an LLM-based generative agent simulator equipped with user profiles, memory, and behavior modules \cite{Zhang:2024aa}. Zhang et al. modeled user-item interactions and their relationships in a recommender system by treating users and items as agents and simulating their interactions through collaborative filtering \cite{Zhang:2024ab}. Guo et al. proposed a framework that positions users and items within a knowledge graph in simulated recommendation scenarios and integrates into the simulation as natural language descriptions \cite{Guo:2024aa}.

\begin{figure}
  \centering
  \includegraphics[width=\linewidth]{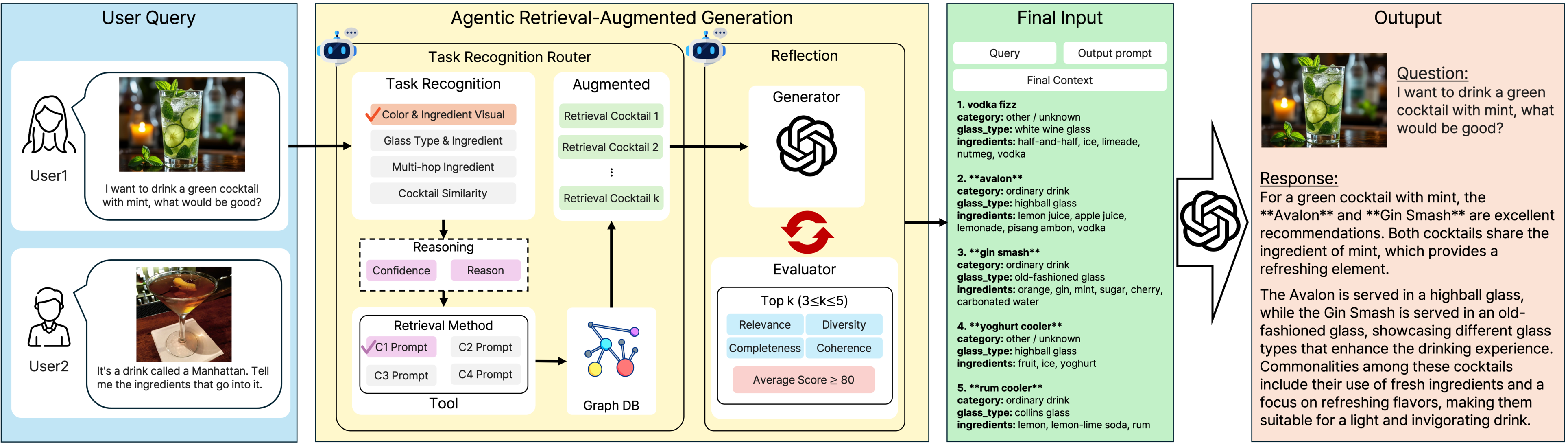}
  \caption{Overall workflow of the MARC}
  \label{MARC}
\end{figure}

\section{Proposed Method}
The overall system proposed in this paper follows the structure and flow of Agentic RAG and is designed for cold-start conditions, where little or no interaction history exists between users and items, rather than for warm-start conditions with abundant interactions. First, we directly constructed a graph database based on cocktail relationships for RAG, which will be used to retrieve context for cocktail recommendations. Subsequently, the system receives the user's cocktail modality information as input, recognizes the task type at the Router stage, and selects the retrieval algorithm. This enables graph database search and selects the top-k candidates. Finally, the Reflection stage sets an appropriate top-k for high-quality recommendations and derives the final context. Both stages are performed by LLMs. Figure \ref{MARC} illustrates the overall workflow of the MARC. The following subsections provide detailed of each component, including graph database construction, the Router, Graph RAG, and Reflection.

\begin{table}[ht]
  \caption{Node, Property, and Relation in Cocktail Knowledge Graph}
  \label{tab:cocktail-kb}
  \centering
  \resizebox{\textwidth}{!}{%
  \begin{tabular}{c p{4cm} p{3.5cm} p{5cm}} 
    \toprule
    Node & Property & \multicolumn{2}{l}{Relation} \\
    \cmidrule(lr){3-4}
         &          & Cypher & Definition \\
    \midrule
    Cocktail   & id, name, name\_embedding, alcoholic, ingredients, drinkThumbnail, ingredientMeasures, description, image description, image description\_embedding, instructions, instructions\_embedding & - & Main Node \\
    Ingredients & name, name\_embedding & (Cocktail)-[:HAS\_INGREDIENT \{measure: ``quantity''\}]->(Ingredient) & Represents the relationship between cocktails and ingredients, storing the quantity of each ingredient using the measure property. \\
    Category   & name, name\_embedding & (Cocktail)-[:CATEGORY]->(Category) & Connects the category to which the cocktail belongs. \\
    GlassType  & name, name\_embedding & (Cocktail)-[:HAS\_GLASSTYPE]->(GlassType) & Indicates the type of glass in which the cocktail is served. \\
    \bottomrule
  \end{tabular}
  } 
\end{table}

\subsection{Graph Database Construction} \label{subsec:GraphDB}
\textbf{Graph Structure.} The Kaggle cocktail dataset is publicly available and features are well-organized, making it easy to build a relational database. Based on this, we collected data including cocktail recipes, ingredients, and metadata. We parsed the ingredient lists and measurements from the raw data, normalized them, and consolidated duplicate ingredient names. For items that were missing or did not exist during this process (e.g., visual image descriptions of cocktails, cocktail explanations), we manually constructed them. Specifically, to express the visual characteristics of each cocktail image as text, we used the Qwen2.5-VL-7B \cite{Bai:2025aa} to generate descriptions. These descriptions include visual elements such as color, texture, and garnish.

A graph database was constructed based on preprocessed data. The graph consists of nodes and relationships, enabling structured information to be effectively retrieved and input as context. Nodes comprise Cocktail, Ingredients, Category, and GlassType. Except for Cocktail, the remaining nodes possess only two properties: name and name\_embedding, which is an embedding of the name property. Relationships between nodes are expressed centered around the main node, Cocktail. The cocktail knowledge graph consists of 4 node types and 3 relation types. The central node, Cocktail, possesses 12 properties. Ingredient, Category, and GlassType nodes connect to it. This graph structure effectively represents the complex relationships within cocktails, enabling multi-hop search and relationship-based recommendations. It particularly facilitates the discovery of new cocktails through common usage patterns among ingredients and supports multi-faceted searches via category and glass type. A detailed explanation of the graph structure can be found in Table \ref{tab:cocktail-kb}.

\noindent \textbf{Vector Representation.} Vector representations are used in various processes during search, including seed node selection, to calculate similarity and select candidates. To achieve this, embedding vectors are generated, and the resulting values are stored as properties of graph nodes. Cocktail, Category, and GlassType were embedded to calculate semantic similarity between item names, while image description was embedded to calculate visual similarity. This enables real-time calculation of cosine similarity between stored embedding values and user queries, allowing for effective retrieval of semantically related cocktails.

\subsection{Agentic RAG: Task Recognition Router, Graph RAG, Reflection} \label{subsec:AgenticRAG}
The main stages of the Agentic RAG used in this study are the Task Recognition Router and Reflection. Task recognition is the Router that determines which search method to use based on the user's question intent. Reflection is the process where LLMs evaluate the quality of search results based on the results themselves, then decide whether to expand the candidate pool by adjusting the top-k value if necessary.

\noindent \textbf{Task Recognition Router.} The Task Recognition Router analyzes the user's natural language query and classifies query into the most suitable task among four defined tasks (C1–C4). Each task employs a search algorithm optimized for a specific aspect of cocktail recommendation and is defined as follows:
\begin{itemize}
\item \textbf{C1 (Color-Ingredient Visual Search):} Processes questions related to color keywords ("red", "blue", etc.) or visual appearance ("elegant", "layered"). The Router detects based on the association between color and ingredients, and visual appeal.
\item \textbf{C2 (Glass Type with Ingredient Matching):} Processes queries specifying a particular glass type ("highball glass", "martini glass", etc.) and considers combinations of glass and ingredients.
\item \textbf{C3 (Multi-hop Ingredient Expansion):} Performs multi-hop graph exploration when only an ingredient list is provided ("cocktails made with whiskey and vermouth").
\item \textbf{C4 (Cocktail Similarity and Alternative):} When a specific cocktail is explicitly targeted ("cocktails similar to Manhattan"), performs relationship-based similarity search.
\end{itemize}
The Router utilizes an LLMs to analyze the intent and keywords of user queries, returning task recognition results in JSON format along with confidence scores and reasons. This enables the automatic selection and execution of the most suitable search algorithm for each query.

\noindent \textbf{Graph RAG.} We implemented graph based retrieval algorithms optimized for each task. Beyond simply using graphs as data storage, the algorithms perform relationship based retrieval by leveraging the structural characteristics and relational information of graph database.
\begin{itemize}
\item \textbf{C1:} Visual feature-based ingredient matching and relationships between cocktails
\item \textbf{C2:} Glass type and ingredient relationships
\item \textbf{C3:} Multi-hop relationships between ingredients and cocktails
\item \textbf{C4:} Ingredient and recipe complexity relationships
\end{itemize}
This effectively satisfies diverse user search intents and enables the discovery of semantically related cocktails. Each recognized task employs a specific retrieval method, with detailed retrieval approaches specified in the Appendix \ref{subsec:appendixA.2}.
	
\noindent \textbf{Reflection.} Reflection is a reflective reasoning process that evaluates the quality of retrieved cocktail candidates and selects the optimal recommendation result. Reflection process is as follows:
\begin{enumerate}
\item \textbf{Multi-Candidate Generation:} Each search algorithm generates top-k candidates. At this stage, different hyperparameter settings (similarity threshold, expansion depth, etc.) are applied to ensure diversity.
\item \textbf{Quality Assessment:} LLMs evaluate each candidate set. The evaluation criteria include the following: Relevance to user query (Relevance score), Diversity of recommendation results (Diversity score), Completeness of information (Completeness score), Consistency of explanation (Coherence score).
\item \textbf{Iterative Refinement:} If the set quality threshold is not met, LLMs gradually expand the top-k value (a search parameter) to perform a re-search (top-k, top-(k+1), top-(k+2), …). This process repeats up to three times, improving the diversity and quality of candidate cocktails. 

\item \textbf{Final Selection:} On iteration if any top-k satisfies the threshold, the process stops immediately and that candidate set is selected as the final output. If no candidate set exceeds the threshold after all iterations, the system selects the set with the highest average score. If there are multiple sets with the same average, the one with the larger k is chosen, so that the model can generate its answer based on richer contextual information.

\end{enumerate}
This Reflection goes beyond simply returning search results, ensuring high-quality recommendations that accurately capture user intent and fit the context. It enables meaningful recommendations even in cold-start conditions by leveraging the relational information within the graph structure.
	
\section{Experiment and Results}
\subsection{Dataset}
The dataset used for constructing the graph database was created by combining Kaggle datasets\footnote{https://www.kaggle.com/datasets/ai-first/cocktail-ingredients, https://www.kaggle.com/datasets/aadyasingh55/cocktails, https://www.kaggle.com/datasets/joakark/cocktail-ingredients-and-instructions} and removing duplicates. Additionally, image descriptions for each cocktail were generated using Qwen-2.5-VL-7B. Description data corresponding to the cocktail descriptions was also manually constructed. Items for which reliable descriptions could not be obtained were excluded, resulting in a final dataset comprising 436 cocktails. 

For evaluation, a separate question dataset consisting of 200 manually designed queries was constructed. These queries were evenly distributed across four task types (50 per task), including 140 multimodal questions and 60 text-only questions. Each task contains 35 items with images, accounting for approximately 70\% of the total items per task. The images used in the experiment were separately constructed to avoid overlap with images in the database.

\subsection{Experiment Configuration}
\noindent \textbf{Implementation Details.} The graph database for Graph RAG was constructed using Neo4j\footnote{https://neo4j.com}, and embeddings were processed using OpenAI's text-embedding-3-small\footnote{https://platform.openai.com/docs/models/text-embedding-3-small}. All LLMs corresponding to each task were processed using GPT-4o-mini.

\noindent \textbf{Metrics.} LLM-as-a-Judge is a method for evaluating and improving LLMs, where LLMs assess the outputs of other models as if they were being evaluated by humans \cite{Zheng:2023aa}. The approach of utilizing LLMs as evaluators in recommender systems was first proposed by Zhang et al. \cite{Zhang:2024ac}, who demonstrated its validity through comparative validation against human evaluations for the same items. While existing evaluation methods have limitations in terms of cost, time, and scalability, LLMs enable efficient and reproducible large-scale evaluations based on their exceptional language understanding and reasoning capabilities. Therefore, this study adopts the LLM-as-a-Judge approach, considering these advantages. The experiment was conducted across four categories: Persuasiveness, Transparency, Accuracy, and Satisfaction, with participants asked to rate each on a scale of 1 to 5 points. Simultaneously, we conduct human evaluation as performed in prior studies, using the same categories and scale for assessment. Detailed descriptions of the evaluation categories and each scale are provided in Appendix \ref{subsec:appendixA.4}.

\begin{table}[ht]
  \caption{LLM-as-a-judge scores by evaluation models and methods. Bold text indicates the highest score in each category.}
  \label{tab:eval}
  \centering
  \resizebox{\textwidth}{!}{%
  \begin{tabular}{c c | c c c c | c}
    \toprule
    Evaluation & Method & Persuasiveness & Transparency & Accuracy & Satisfaction & Average \\
    \midrule
    \multirow{2}{*}{GPT-4o-mini\footnotemark[4]} 
     & w/o Graph, Reflection & 3.64 & 4.45 & 3.41 & 3.63 & 3.78 \\
      & w/o Reflection & 4.06 & \textbf{4.71} & \textbf{3.73} & 4.06 & 4.14 \\
      & w/o Graph & 3.84 & 4.57 & 3.59 & 3.85 & 3.96 \\
      & MARC (Ours) & \textbf{4.12} & 4.70 & \textbf{3.73} & \textbf{4.11} & \textbf{4.17} \\
    \midrule
    \multirow{2}{*}{GPT-5\footnotemark[4]} 
      & w/o Graph, Reflection & 1.93 & 2.84 & 1.73 & 1.69 & 2.05 \\
      & w/o Reflection & 2.16 & \textbf{3.01} & \textbf{1.99} & 1.92 & \textbf{2.27} \\
      & w/o Graph  & 1.97 & 2.82 & 1.81 & 1.78 & 2.09 \\
      & MARC (Ours)  & \textbf{2.19} & 2.97 & 1.98 & \textbf{1.93} & \textbf{2.27} \\
    \midrule
    \multirow{2}{*}{Human} 
      &  w/o Graph, Reflection & 4.16 & 4.21 & 4.13 & 3.82 & 4.08 \\
      &  w/o Reflection &  \textbf{4.39} &  \textbf{4.47} & 4.28 & 4.17 & 4.33 \\
      &  w/o Graph & 4.36 & 4.40 & 4.37 & 4.22 & 4.34 \\
      & MARC (Ours) & 4.35 & 4.46 &  \textbf{4.45} &  \textbf{4.26} & \textbf{4.38} \\      
    \bottomrule
  \end{tabular}
  }
\end{table}
\footnotetext[4]{https://openai.com/api/}

\subsection{Results}
The quantitative performance of MARC is reported using two evaluation metrics. First, we present the results of the LLM-as-a-Judge based automated evaluation, followed by the results of the Human Evaluation conducted using the same metrics. We performed ablation tests on 200 identical responses generated by GPT-4o-mini, comparing MARC (Ours), vector RAG only (w/o Graph), without Reflection (w/o Reflection), and vector RAG only and without Reflection (w/o Graph, Reflection) approaches. To ensure fairness, 13 questions that yielded no search results due to database limitations were excluded. For the remaining 187 questions, we performed LLM-as-a-Judge and the evaluation models used were GPT-4o-mini and GPT-5, cross-validated against each other.

For LLM-as-a-Judge in both evaluation models, MARC achieved the highest overall average score. In the GPT-4o-mini evaluation, MARC achieved an average of 4.17 points, while other approaches scored 3.78, 3.96, and 4.14 points, showing relative improvements ranging from +0.39 to +0.03. The GPT-5 evaluation also showed a relative improvement, with MARC recording an average of 2.27 points compared to other approaches averaging 2.05, 2.09, and 2.27 points, corresponding to gains of +0.22 to +0.00. Comparing the evaluation models, GPT-5 was relatively conservative. This result is consistent with the findings of Abdoli et al. \cite{Abdoli:2025aa}, which reported that GPT-5 tends to assign conservative scores and shows higher variability compared to other GPT versions. Nevertheless, the fact that the Transparency metric consistently outperformed GPT-5 results suggests that the graph structure more clearly reveals the causal pathways underlying recommendations, thereby enhancing the traceability of explanations (why-explanation clarity).

\begin{figure}
  \centering
  \includegraphics[width=\linewidth]{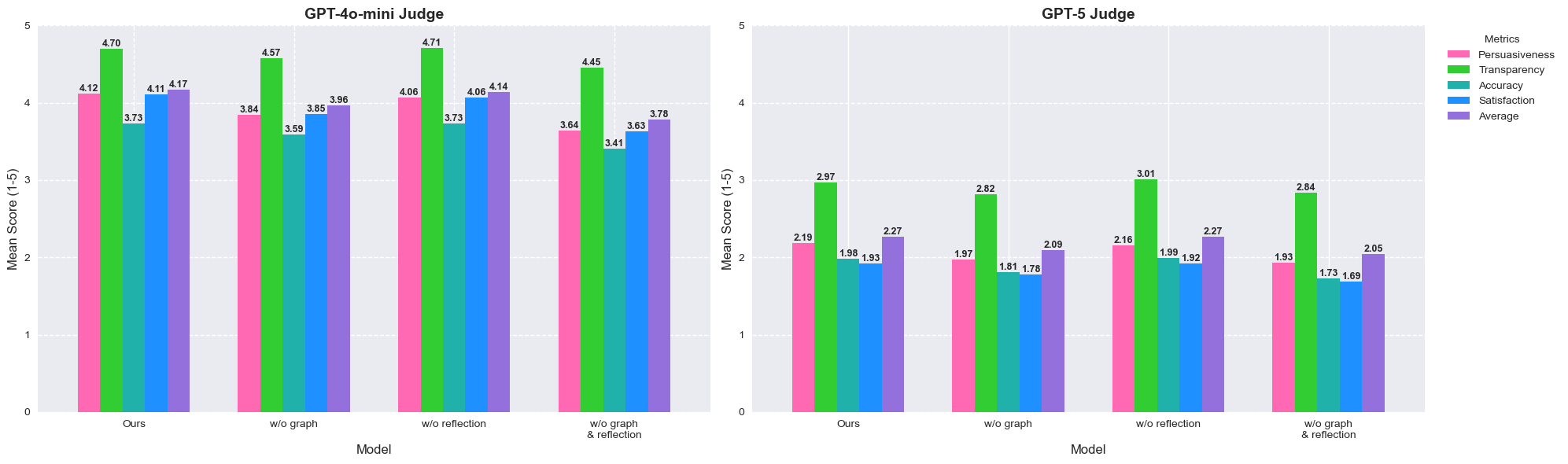}
  \caption{Visualization of Evaluation Results using LLM-as-a-Judge (Left: GPT-4o-mini, Right: GPT-5)}
\end{figure}

Human evaluation was conducted by four individuals. A total of 80 samples were randomly shuffled and presented, consisting of 10 questions per task extracted for all approaches. Similar to LLM-as-a-Judge, three cases where search results were unavailable due to database limitations were excluded, resulting in a final evaluation of 154 questions. Human evaluation results showed MARC achieved an average of 4.38 points, while the other approaches scored 3.78, 3.96, and 4.14 points, showing relative improvements ranging from +0.30 to +0.04. In the human evaluation, while the overall average and metrics such as Transparency, Accuracy, and Satisfaction generally increased, Persuasiveness showed a slight decline compared to w/o Reflection and w/o Graph. This indicates that non-structural elements such as writing style, tone, and summary density influence the impression of the response on actual humans, suggesting the need for additional calibration. 

We compared all the approaches, MARC was the best overall in all the evaluation models. However, the results suggest that the Graph RAG and database contributed most significantly to performance improvement. The incorporation of the Reflection further enhanced overall quality but showed marginal or inconsistent effects in certain metrics. This tendency can be interpreted from two perspectives. First, regarding top-k selection, the hyperparameter was fixed at 3 (tested within the range of 3–5), which ensures stable performance but may have limited the potential benefit of the Reflection loop. If top-k = 1 had been applied, a stronger positive effect of Reflection could likely have been observed, since the Reflection would have played a more decisive role in expanding candidate diversity.Second, the threshold for response quality was also set empirically at 80 and was not systematically tuned. A stricter threshold could have provided more refined filtering and thus higher-quality responses. Therefore, while both modules jointly contribute to performance, the graph structure is the dominant factor driving improvement, and optimizing Reflection-related hyperparameters remains an important direction for future work.

\section{Conclusion}
This study proposes MARC for cold-start by constructing a graph database based on cocktail data. The Task Recognition Router formalizes user queries into tasks, and responses are generated based on the threshold set in Reflection. Results from LLM-as-a-Judge and human evaluation show that our approach achieved overall balanced improvements. Although the human evaluation revealed a slight decrease in Persuasiveness, implying that linguistic style and prompt formulation still influence how users perceive the responses. Addressing these linguistic and stylistic factors will be an important direction for future prompt refinement. Limitations include: applied only to cocktail data, being restricted to cold-start conditions, utilizing only a graph database, employing a non-refined threshold, and relying on text-based similarity without image embeddings. Future research will focus on other domains (e.g. movies), warm-start conditions, implementing a Multi-agent-based RAG and reflection, and developing an end-to-end system using image embeddings.

\section*{Declaration on Generative AI}
During the preparation of this work, the author(s) used DeepL and ChatGPT for translation, grammar, and spelling check. After using this tool/service, the author(s) reviewed and edited the content as needed and take(s) full responsibility for the publication’s content.

\bibliography{mmagent_recsys_ref}

\appendix
\section{Appendix}

\subsection{Specific Graph Retrieval  Algorithm} \label{subsec:appendixA.2}
\textbf{C1: Color-Ingredient Visual Search Algorithm.} C1 perform hybrid search combining graph structure and embeddings, focusing on the visual characteristics of cocktails. We extract the cocktail name, ingredients, category, and color from the user query. Next, it searches for ingredient nodes semantically associated with color keywords extracted from the user query (e.g., "red", "orange"). It calculates the similarity between the color keyword embedding and the ingredient name embedding to identify the most relevant ingredient, then locate cocktails that are jointly referenced by each attributes. For each candidate cocktail, we calculate the cosine similarity between its image description embedding and the user query embedding. Finally, all cocktails are ranked from highest to lowest similarity, and the top-k results are selected as the final recommendations. By leveraging both ingredient and image description information, we recommend cocktails that best match the user's visual intent.

\noindent \textbf{C2: Glass Type-Ingredient Matching.} C2 explore related cocktails centered around the glass type and ingredients specified by the user. From user queries explicitly mentioning a glass type, extract the GlassType, Cocktail, Ingredients, and Category. First, filter Cocktail nodes matching the extracted GlassType. Then perform Ingredient matching. At this step, search for cocktails matching all ingredients. If insufficient results are found, gradually relax the ingredient condition by excluding one ingredient at a time, starting with the last ingredient extracted from the user query. This iteration continues until k candidate cocktails are secured. Finally, the final cocktail is selected by sorting the similarity scores between the extracted cocktail's image description embedding and the user query in descending order. This approach provides results aligned with the user intent to explore cocktails with similar ingredients within the same glass type. 

\noindent \textbf{C3: Multi-hop Ingredient Expansion Search.} C3 discovers new cocktails across up to 3 hops by exploring ingredient relationships through multi-hop graph traversal. First, it extracts cocktails and ingredients from the user query. At the 1-hop level, it searches for Cocktail nodes directly connected to the input ingredient via a HAS\_INGREDIENT relationship. Only cocktails containing at least n matching ingredients are selected. At the 2-hop level, it analyzes ingredient patterns common to cocktails discovered in the first hop. For example, it uncovers hidden association patterns between ingredients, such as "Cocktails using mint and lime also commonly use rum." Only cocktails sharing at least n common ingredients proceed to the next step. At the 3-hop level, it explores new cocktails based on the discovered common ingredients. For each new cocktail found at the 3-hop level, it calculates the following two scores:
\begin{itemize}
  \item \textbf{Extension strength:} How many of the common ingredients found in 2-hop are included in this cocktail?
  \item \textbf{Ingredient bonus:} How many of the original ingredients entered by the user are included in this cocktail?
\end{itemize}
The total score obtained by summing the counts from these two components is used to sort the results in descending order. In case of a tie, cocktails with higher expansion strength are ranked higher. The candidate cocktails selected through this multi-hop expansion are finally ranked in descending order by calculating the cosine similarity between the user query and the image description embedding of each cocktail. This multi-hop search recommends cocktails related by ingredients that are difficult to discover through direct ingredient matching. It uses cosine similarity to recommend cocktails, considering both the structural characteristics of the graph and semantic similarity.

\noindent \textbf{C4: Material-based Similar Recipe Cocktail Recommendation.} C4 comprehensively analyzes graph-based similarity and ingredient complexity to recommend alternative cocktail recipes. First, it extracts the target cocktail and ingredients from the user query as seeds. Prioritize exploring Cocktail nodes matching the seed cocktail extracted from the database. If the seed cocktail cannot be found in the user query, embed the entire query and return cocktails with high cosine similarity to the name\_embedding as potential seeds. Then, explore other cocktails sharing ingredients identical to those in the seed cocktail. For each cocktail, the number of ingredients shared with the seed cocktail is calculated, while also determining the total number of ingredients in that cocktail. During the complexity analysis phase, only cocktails where the difference in ingredient count between the seed cocktail and the candidate cocktail is within m are selected. Additionally, only cocktails sharing at least x ingredients are retained. The final ranking is sorted in descending order based on the number of shared ingredients. In case of a tie, the top-k simpler cocktails with fewer ingredients are selected. This graph relationship enables the recommendation of alternative cocktails with similar ingredient compositions and appropriate complexity. C4 focuses on structural similarity rather than semantic similarity, providing practical recommendations that consider actual manufacturability and taste similarity.

\subsection{Task-specific Prompt Template}
This section presents the original text of the prompt used in the framework of this study. The prompts for each process and task within the framework are written in English, accompanied by brief descriptions of each template. Only the Reflection prompt is attached below and detailed prompts are available on the publicly released GitHub repository. 

\lstset{
  basicstyle=\ttfamily\scriptsize,       
  breaklines=true,                       
  frame=single,                          
  backgroundcolor=\color{gray!10},       
  keywordstyle=\color{black},            
  commentstyle=\color{green!60!orange},  
  stringstyle=\color{black},             
  showstringspaces=false
}
	
\begin{lstlisting}[language=Python, caption={Reflection Prompt Template}, basicstyle=\ttfamily\scriptsize]
REFLECTION_PROMPT_TEMPLATE = """You are a quality evaluation expert for a cocktail recommender system.
Based on the user's query and the retrieved cocktail candidates, evaluate the quality according to the following four criteria.

## User Query
{user_query}

## Retrieved Cocktail Candidates ({num_results} items)
{search_results}

## Evaluation Criteria
1. **Relevance**: How relevant are the search results to the user's query? (0-100)
   - Do they reflect the user's desired characteristics (color, flavor, ingredients, style, etc.)?
   - Do they satisfy the core requirements of the query?

2. **Diversity**: How diverse are the recommended cocktails? (0-100)
   - Do they offer different styles, flavors, and ingredient combinations?
   - Do they provide a broad range of options rather than monotonous results?

3. **Completeness**: How comprehensive are the recommendations? (0-100)
   - Do they sufficiently provide the information the user is seeking?
   - Is there a possibility that better alternatives are missing?

4. **Coherence**: How logically consistent are the recommendations? (0-100)
   - Are the reasons for recommendation clear and valid?
   - Do the recommendations form a harmonious set overall?

## Output Format (JSON)
{
    "relevance": 85,
    "diversity": 70,
    "completeness": 80,
    "coherence": 90,
    "overall_score": 81.25,
    "feedback": "High relevance but low diversity. Many cocktails have similar styles; providing more varied options would be better.",
    "suggestions": [
        "Increase diversity in color or base spirits",
        "Consider adding alcoholic/non-alcoholic options"
    ],
    "should_retry": False
}

Each score should be between 0 and 100, with overall_score as the average of the four scores.
Set should_retry to true if the overall score is below 80.
"""
\end{lstlisting}

\noindent \textbf{Reflection Template.} This template performs reflection based on generated responses to expand the number of retrieved cocktails and enhance context. It is the second core step proposed in this study. If the set threshold is not exceeded, the process repeats until it is, ensuring high-quality information retrieval and enabling input into LLMs' context.

\subsection{Detailed Description of Evaluation Metrics} \label{subsec:appendixA.4}
The experiment was conducted using the following items:
\begin{itemize}
  \item \textbf{Persuasiveness:} “This explanation is convincing to me.”
  \item \textbf{Transparency:} “Based on this explanation, I understand why this movie is recommended.”
  \item \textbf{Accuracy:} "This explanation is consistent with my interests.”
  \item \textbf{Satisfaction:} “I am satisfied with this explanation.”
\end{itemize}
For each item, LLMs instructed to evaluate on a scale of 1 to 5 points. 1 point was set as Strongly Disagree, 2 points as Disagree, 3 points as Neutral, 4 points as Agree, and 5 points as Strongly Agree.

\end{document}